\documentclass[aps,prl,twocolumn,10pt,nobibnotes]{revtex4-2}
\usepackage{amsmath,amssymb,graphicx,mathtools,dsfont}
\usepackage[T1]{fontenc}
\usepackage[dvipsnames]{xcolor}

\begin{document}

\newcommand{\bra}[1]    {\langle #1|}
\newcommand{\ket}[1]    {|#1 \rangle}
\newcommand{\ketbra}[2]{|#1\rangle\!\langle#2|}
\newcommand{\braket}[2]{\langle#1|#2\rangle}
\newcommand{\tr}[1]    {{\rm Tr}[ #1 ]}
\newcommand{\titr}[1]    {\widetilde{{\rm Tr}}\left[ #1 \right]}
\newcommand{\avs}[1]    {\langle #1 \rangle}
\newcommand{\modsq}[1]    {| #1 |^2}
\newcommand{\0}    {\ket{\vec 0}}
\newcommand{\1}    {\ket{\vec 1}}
\newcommand{\dt}    {\delta\theta}
\newcommand{\I}    {\mathcal  I_{rc}^{(q)}}
\newcommand{\Ic}    {\mathcal  I_{rc}}
\newcommand{\C}    {\hat{\mathcal C}_i(t)}
\newcommand{\Cd}    {\hat{\mathcal C}_i^\dagger(t)}
\newcommand{\trr}[1]    {{\rm Tr}_{rc}[ #1 ]}

\title{Limits to velocity of signal propagation in many-body systems:\\
  a quantum-information perspective}
\author{Piotr Wysocki and Jan Chwede\'nczuk}
\affiliation{Faculty of Physics, University of Warsaw, ul. Pasteura 5, 02-093 Warszawa, Poland}
\begin{abstract}
  The Lieb-Robinson bound (LRB) states that the range and strength of interactions between the constituents of a complex many-body system impose upper limits to how fast the signal can propagate. 
  It manifests in a light cone-like growth of correlation function connecting two distant subsystems. 
  Here we employ the techniques of quantum information
  to demonstrate that the LRB can be determined from local measurements performed on a single qubit that is connected to a many-body system. 
  This formulation provides an operational recipe for estimating the LRB in complex systems, replacing the measurement of the correlation function with simple single-particle manipulations. 
  We demonstrate the potency of this approach by deriving
  the upper limit to the speed of signal propagation in the XY spin chain.
\end{abstract}

\maketitle

The Lieb-Robinson bound imposes an upper limit to the speed of signal propagation in quantum systems~\cite{lieb1972finite}. It is
quantified by considering two systems, $A$ and $B$, distanced by $d$ and evolving according to a Hamiltonian $\hat H$. The LRB states that the norm of the commutator 
$[\hat A,\hat B(t)]$ is bounded by
\begin{align}\label{eq.lrb}
  \|[\hat A,\hat B(t)]\|\leqslant ae^{-(d-vt)},
\end{align}
where $a$ depends on the operators $\hat A$, $\hat B(t)$ and the Hamiltonian, while the propagation speed $v$ is determined solely by the details  of $\hat H$. 
In the wake of the discovery of this fundamental relation, vast experimental and theoretical effort was invested into 
measuring or calculating this bound for various systems. The LRB is very important for the development of quantum technologies, as it governs the information spreading across quantum circuits~\cite{PhysRevA.81.062107,PhysRevLett.117.091602,PhysRevLett.127.160401,PhysRevLett.97.050401,annurev:/content/journals/10.1146/annurev-conmatphys-031720-030658,gebert2016polynomial,PhysRevA.101.022333,PhysRevA.100.032311}, and
its role has been recognized as a limiting factor to the performance of quantum heat engines~\cite{PhysRevE.96.022138}.
It is also an important probe of the microscopic properties of many-body quantum systems~\cite{PRXQuantum.1.010303,PhysRevB.105.125101,PhysRevLett.104.190401}. 
It is a determinant of correlations strength build-up in complex systems~\cite{PhysRevLett.111.230404,PhysRevLett.111.207202,nachtergaele2006propagation,faupin2022lieb,mahoney2024lieb}, a quantity that is
directly measurable in experiment~\cite{cheneau2012light}. Recently it has been shown that the LRB influences the rate of entanglement growth
in many-body systems~\cite{PhysRevLett.130.170801,PhysRevLett.132.100803}.

Usually, the LRB is obtained from the measurement of the correlation function between the two distant subsystems. We propose a different route, and show that 
the LRB can be determined with simple local measurements on a single qubit that is connected to a many-body system. This stems from the central observation of this work---the amount of
information about the initial perturbation applied to some part of the system distant from the qubit, which propagates under a Hamiltonian $\hat H$, is
inherently bounded by the LRB. This result is obtained using the tool known as the quantum Fisher information (QFI)~\cite{braunstein1994statistical}. Although it is usually invoked in the
context of quantum metrology, as it sets the lower bound to the sensitivity of quantum sensors~\cite{pezze2009entanglement}, the QFI is a versatile tool, allowing to
determine the strength of multipartite entanglement~\cite{hyllus2012fisher,toth2012multipartite,hauke2016measuring}, the Einstein-Podolsky-Rosen steering~\cite{yadin2021metrological}, or the Bell nonlocality~\cite{PhysRevLett.126.210506}. It is also a fine probe of quantum- and thermal-phase transitions 
in complex systems~\cite{PhysRevE.84.041116,wang2014quantum,PhysRevB.96.104402,PhysRevA.80.012318,PhysRevA.78.042106,PhysRevA.78.042105,PhysRevA.90.022111,
  PhysRevE.93.052118,PhysRevLett.124.120504,PhysRevLett.123.170604,hauke2016measuring}. Here, the QFI again proves to be a potent tool, opening a way towards a simpler determination of the LRB in many-body
systems and shedding new light on the relation between various many-body effects and information propagation~\cite{https://doi.org/10.1002/andp.201600280,PhysRevB.99.054204,PhysRevB.99.064303}.

\begin{figure}[t!]
    \centering
    \includegraphics[width=0.9\linewidth]{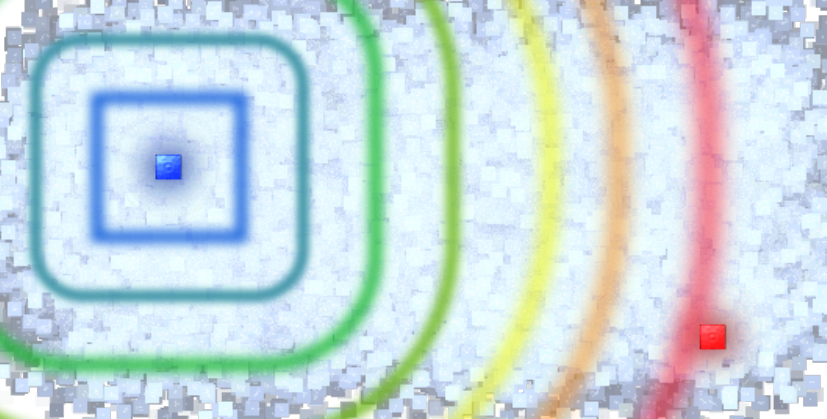}
    \caption{Illustrarion showing the source (dark blue square) sending the signal through the many-body system to the single receiving qubit. The LRB is derived from the single-particle local
    operations.}
    \label{fig.scheme}
\end{figure}

The process, illustrated in Fig.~\ref{fig.scheme}, 
starts with an impulse that acts for time $\tau$ on some subsystem that becomes a {\it source} of the signal. This first step can be expressed as
\begin{align}\label{eq.singal}
  \hat\varrho(\theta)=e^{-i\theta\hat H_{sr}}\hat\varrho\, e^{i\theta\hat H_{sr}},
\end{align}
where $\hat\varrho$ is the initial density operator of the full system and $\hat H_{sr}$ is the (local) source Hamiltonian. The parameter $\theta=\omega\tau$ is the measure of the 
coupling strength to the external potential and $\omega$ is some characteristic frequency.

Subsequently, a Hamiltonian $\hat H$ triggers the propagation of the signal across the system, yielding the outcome
\begin{align}\label{eq.evo.sr}
  \hat\varrho(\theta;t)=e^{-it\hat H}\hat\varrho(\theta)\, e^{it\hat H}.
\end{align}
The main question of this work can be stated as follows: how fast does the information about $\theta$, the parameter connected to the triggering impulse, 
propagate to the {\it receiver}---a subsystem distant from the source?

We provide an answer making a single assumption: the receiver is a single qubit with its local density operator
\begin{align}\label{eq.trace.til}
  \hat\varrho_{rc}^{(\theta)}(t)=\titr{\hat\varrho(\theta;t)},
\end{align}
where the tilde denotes the partial trace over all the non-receiver dergees of freedom.

The information about the impulse that can be extracted from any measurements on the receiver is no larger than the QFI, which is denoted by $\I$ and has a
general form of
\begin{align}\label{eq.qfi}
  \I=2\sum_{i,j}\frac{\modsq{\bra{\psi_j}\dot{\hat\varrho}_{rc}^{(\theta)}(t)\ket{\psi_i}}}{\lambda_i+\lambda_j},
\end{align}
where the summation runs over the elements of the spectrum of $\hat\varrho_{rc}^{(\theta)}(t)$, $\ket{\psi_{i/j}}$ are its eigenstates and $\lambda_{i/j}$ the corresponding eigenvalues.
The dot denotes the derivative of the density operator with respect to the parameter $\theta$. For a receiver that is a qubit, Eq.~\eqref{eq.qfi} reduces to
\begin{align}
  \I=\sum_{i=1}^3\modsq{\tr{\hat\sigma_i\dot{\hat\varrho}(\theta,t)}}.
\end{align}
Here $\hat\sigma_{j}=\lambda_j^{-1/2}\ketbra{\psi_{j}}{\psi_{j}}$ ($j=1,2$) and $\hat\sigma_3=\hat\sigma_{rc}^{(x)}\cos\phi+\hat\sigma_{rc}^{(y)}\sin\phi$, where 
$\phi$ is the phase of the matrix element $\bra{\psi_1}\dot{\hat\varrho}_{rc}^{(\theta)}(t)\ket{\psi_2}$ (see Appendix).  
The trace now runs over the Hilbert space of the whole system. The $\I$ is a convex functional of the density operator~\cite{pezze2009entanglement}, hence
by using Eq.~\eqref{eq.singal} and the spectral decomposition of the density matrix $\hat\varrho(\theta)=\sum_np_n\ketbra{\Psi_n}{\Psi_n}$, we obtain
\begin{align}\label{eq.conv.2}
  \I&=\sum_{i=1}^3\modsq{\tr{\hat\sigma_i e^{-it\hat H}[\hat H_{sr},\sum_np_n\ketbra{\Psi_n}{\Psi_n}]e^{it\hat H}}}\nonumber\\
  &\leqslant\sum_{i=1}^3\sum_np_n\modsq{\tr{\hat\sigma_i(t)[\hat H_{sr},\ketbra{\Psi_n}{\Psi_n}]}},
\end{align}
where $\hat\sigma_i(t)=e^{it\hat H}\hat\sigma_i e^{-it\hat H}$. The trace can be calculated using the basis of $\ket{\Psi_n}$'s, giving
\begin{align}\label{eq.ineq.i}
  \I\leqslant\sum_{i=1}^3\sum_np_n\modsq{\bra{\Psi_n}[\hat\sigma_i(t),\hat H_{sr}]\ket{\Psi_n}}.
\end{align}
The emergent commutators $\C\equiv[\hat\sigma_i(t),\hat H_{sr}]$ of operators related to the source and receiver (at time $t$)
are the core of the original Lieb-Robinson analysis~\cite{lieb1972finite}, see Eq.~\eqref{eq.lrb}. The final step is to note that
\begin{align}\label{eq.ineq.double}
  &\modsq{\bra{\Psi_n}\C\ket{\Psi_n}}=\bra{\Psi_n}\Cd(\hat{\mathds1}-\hat\Pi_n^\perp)\C\ket{\Psi_n}\nonumber\\
  &\leqslant\bra{\Psi_n}\Cd\C\ket{\Psi_n}\leqslant\|\C\|^2,
\end{align}
where $\hat\Pi_n^\perp$ projects onto the subspace orthogonal to that spanned by $\ketbra{\Psi_n}{\Psi_n}$ and $\|\C\|^2$ is the operator norm. Substitution of this inequality into line~\eqref{eq.ineq.i}
gives
\begin{align}\label{eq.final}
  \I\leqslant\sum_{i=1}^3\|\C\|^2.
\end{align}
This is the central result of this work---the amount of information that can be extracted from the receiver is upper-bounded by the LRB. 
To estimate or measure the LRB it is thus sufficient to perform local single-body measurements. To recapitulate, this derivation does not rely on any assumptions about the 
input state $\hat\varrho$, nor about the system, or the Hamiltonian $\hat H$. Only the initial transformation is assumed to be in the general form of Eq.~\eqref{eq.singal}
and the receiver to be a single qubit. 

We now discuss the conditions under which this inequality can be saturated. The line~\eqref{eq.conv.2} becomes an equality when the state $\hat\varrho(\theta)$ is pure. The first inequality 
in line~\eqref{eq.ineq.double} is saturated when $\ket{\Psi_n}$ is an eigenstate of $\C$ and the last relation is saturated when the modulus of its eigenvalue is maximal.

Finding the spectrum of $\C$'s is usually hard. We show that the relation between the QFI and the LRB not only sheds some new light on both these quantities, but also allows to
analytically derive the LRB using the knowledge of $\I$. For illustration, we pick the XY spin chain depicted by the Hamiltonian
\begin{align}\label{eq.ham.xy}
  \hat H=\sum_j(\hat\sigma_j^{(x)}\hat\sigma_{j+1}^{(x)}+\hat\sigma_j^{(y)}\hat\sigma_{j+1}^{(y)}),
\end{align}
which is an excellent model for analysing the propagation of the signal through a system~\cite{PhysRevA.100.052309}. 
The impulse acts on the source qubit, and its generator is one of the Pauli matrices, say $\hat\sigma^{(x)}_{sr}$.
Subsequently, the Hamiltonian~\eqref{eq.ham.xy} propagates the signal to the receiver, positioned $n$ qubits away from the source.

\begin{figure}[t!]
    \centering
    \includegraphics[width=0.9\linewidth]{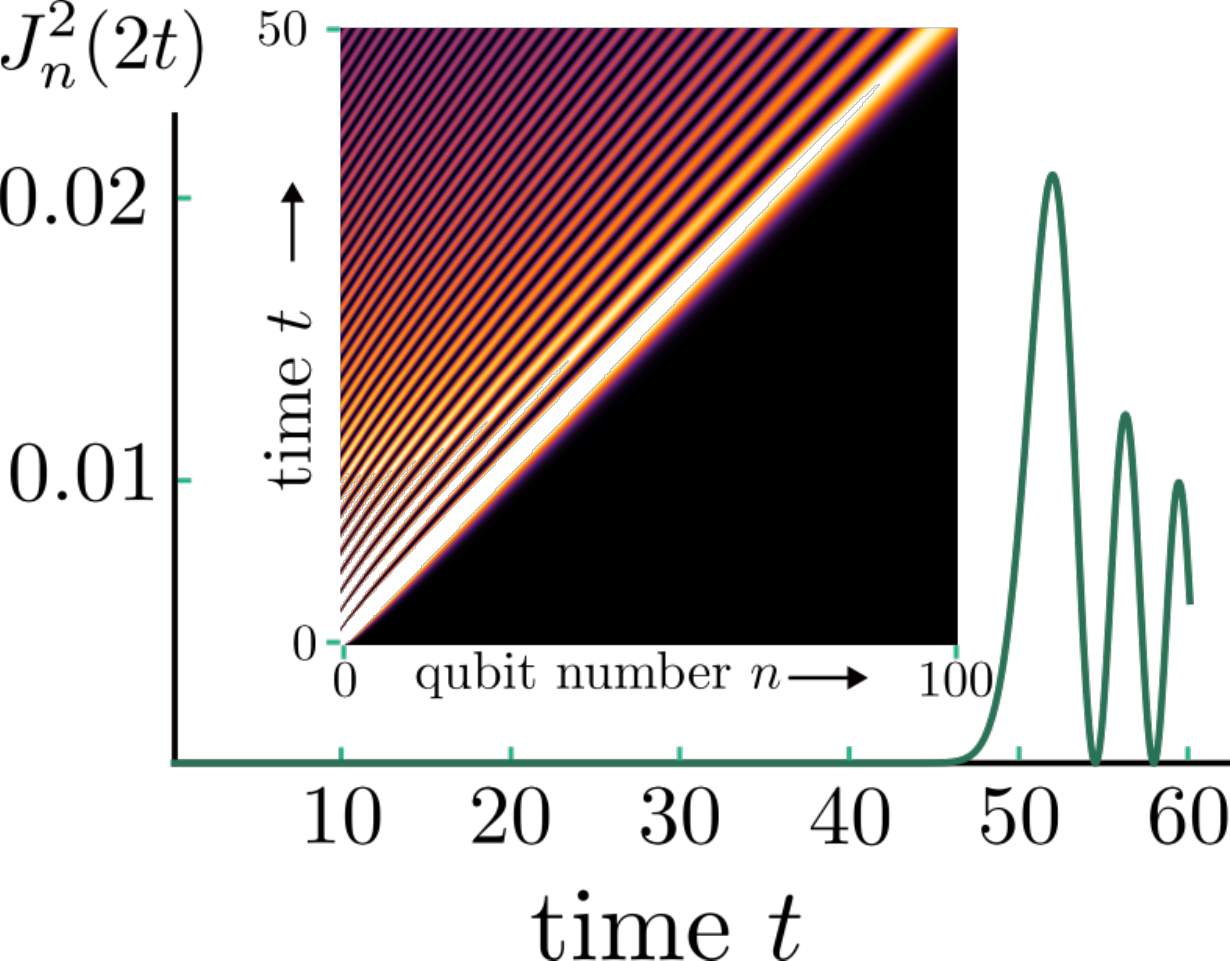}
    \caption{The main figure shows the time-dependence of the Bessel function $J^2_n(2t)$ for $n=100$. The inset is a space-time plot of $J^2_n(2t)$ ($n\in[0,100]$ and $t\in[0.50]$), showing the characteristic 
      light cone-like structure.}
    \label{fig.cone}
\end{figure}

In order to saturate Eq.~\eqref{eq.final} we shall determine the maximal value of $\I$. According to~\cite{braunstein1994statistical},
\begin{align}\label{eq.qfi.bound}
  \I\leqslant\sum_k\frac1{p_k}\left(\frac{\partial p_k}{\partial\theta}\right)^2+4\Delta^2\hat h,
\end{align}
where $\Delta^2\hat h=\avs{\hat h^2}-\avs{\hat h}^2$,  $\hat h$ generates the infinitesimal  transformation of 
$\hat\varrho_{rc}^{(\theta)}(t)$ to $\hat\varrho_{rc}^{(\theta+\dt)}(t)$ and $p_k$ are the eigen-values of $\hat\varrho_{rc}^{(\theta)}(t)$. The above inequality is saturated if 
$\hat\varrho_{rc}^{(\theta)}(t)=\ketbra{\psi_{rc}^{(\theta)}(t)}{\psi_{rc}^{(\theta)}(t)}$, i.e., it is pure, and this happens
when the input state $\hat\varrho=\ketbra{\psi}{\psi}$ is the density operator representing, symmetrically, either of the two pure states 
\begin{align}\label{eq.states}
  \ket{\psi}=\ket{0,\ldots,0}\equiv\0,\ \ \ \mathrm{or}\ \ \ \ket{\psi}=\ket{1,\ldots,1}\equiv\1,
\end{align}
where $\ket0$ and $\ket1$ are the single-qubit eigenstates of $\hat\sigma^{(z)}$. These are the only states of zero eigenvalue of the Hamitonian~\eqref{eq.ham.xy}, yielding a non-entangled
state at any time $t$. Other states will entangle, producing a mixed output at the receiver. We take one of the states from~\eqref{eq.states}, for instance $\ket{\psi}=\0$ as the input state and choose the
working point $\theta=0$~\footnote{For all other $\theta$'s the input state must be of the form $\ket\psi=e^{i\theta h}\0$ or $\ket\psi=e^{i\theta h}\1$ to accommodate for the shift of 
the working point}, giving, up to the dominant order (see Appendix),
\begin{align}\label{eq.pure.rc}
  \ket{\psi_{rc}^{(\theta)}(t)}=(\hat{\mathds1}-i\theta J_n(2t)\hat\sigma_{rc}^{(x)})\ket0_{rc},
\end{align}
where $J_n(2t)$ is the Bessel function of the first kind. Note that since for small $\theta$ the input state $\0$ is trasformed into another pure state~\eqref{eq.pure.rc}, the first term of Eq.~\eqref{eq.qfi.bound} does not contribute to $\I$, up to the leading order in $\theta$. Hence the generator of the transformation is $J_n(2t)\hat\sigma_{rc}^{(x)}$, and plugged into Eq.~\eqref{eq.qfi.bound} gives 
\begin{align}\label{eq.qfi.max}
  \I\leqslant 4J^2_n(2t),
\end{align}
which, according to Eq.~\eqref{eq.final}, provides the LRB for this model. Note that once the relation~\eqref{eq.final} was established, the derivation of the LRB required only the knowledge of
the properties of $\I$. The Fig.~\ref{fig.cone} shows the dependence of $J_n^2(2t)$ on time for $n=100$. The signal grows exponentially around $t=n/2=50$. Varying both $n$ from 0 to 100 and $t$ from 0
to 50 highlights the emblematic light cone-like structure with the speed of propagation, see Eq.~\eqref{eq.lrb}, equal to $v=2$.

We note that the QFI from Eq.~\eqref{eq.qfi} is the result of maximization of the Fisher information
\begin{align}\label{eq.cfi}
  \Ic=\sum_{x}\frac1{p(x|\theta)}\left(\frac{\partial p(x|\theta)}{\partial\theta}\right)^2
\end{align}
over all possible measurement operators $\hat E_{rc}(x)$~\cite{braunstein1994statistical}, where the probability of obtaining a result $x$ is
\begin{align}
  p(x|\theta)=\tr{\hat\varrho_{rc}^{(\theta)}(t)\hat E_{rc}(x)}
\end{align}
and $\sum_x\hat E_{rc}(x)=\hat{\mathds1}$, $\hat E_{rc}(x)\geqslant0$. 
While in general it is hard to determine the set of $\hat E_{rc}(x)$'s for which the Fisher information from Eq.~\eqref{eq.cfi} saturates the QFI from Eq.~\eqref{eq.qfi},
in the case of the Hamiltonian~\eqref{eq.ham.xy}, according to Eq.~\eqref{eq.pure.rc} the probabilities of finding the receiver qubit in states $\ket0_{rc}$ and $\ket1_{rc}$ around the working point $\theta=0$,
up to the leading order, are $p(0|\theta)\simeq(1-\theta^2J^2_n(2t))$ and $p(1|\theta)\simeq\theta^2J^2_n(2t)$, which plugged into Eq.~\eqref{eq.cfi} gives $\Ic=4J^2_n(2t)$. Hence the simple single-qubit measurements allow to saturate the 
QFI bound and in consequence determine the LRB. 

It is worth noting that the steps that are necessary to estimate the LRB
have already been demonstrated experimentally in a different context.
The small (quasi-infinitesimal) increments of $\theta$ were implemented experimentally in~\cite{hyllus2012fisher} for a similar purpose---to measure
local properties of a probability and to estimate the value of the Fisher information a complex many-body system. 

In conclusion, we have demonstrated that the Lieb-Robinson bound can be derived from local measurements on a single qubit connected to a many-body system, hence shifting the necessity 
of measuring correlation functions between distant subsystems.
This is possible, because as we have shown, the LRB intrinsically limits the amount of information that reaches a single qubit that is a part of a complex many-body system.
We have identified the conditions under which the local measurements on this qubit can saturate the upper limit set by the LRB, hence allowing one to determine the LRB from simple one-body measurements.
We have used an example of the XY spin chain to show how the two formulations---that of the LRB and of quantum information---interplay, allowing to determine the velocity of information propagation. 
Due to the relative simplicity of the one-qubit operations and measurements, the protocol presented in this work might find applications in future measurements of the LRB, hence
contributing to our understanding of complex many-body systems.
Finally, we underline that one could independently derive Eq.~\eqref{eq.final} by starting from the analysis of information propagation through many-body quantum systems~\cite{PhysRevLett.97.050401}.

We acknowledge the discussion with Mi{\l}osz Panfil and Tomasz Wasak. We thank Iman Sargolzahi for a critical reading of our paper.
This work was supported by the National Science Centre, Poland, within the QuantERA II Programme that has received funding from the European Union’s Horizon 2020 
research and innovation programme under Grant Agreement No 101017733, Project No. 2021/03/Y/ST2/00195.

\clearpage
\onecolumngrid

\appendix
\setcounter{equation}{0}
\renewcommand{\theequation}{A.\arabic{equation}}

\section{Derivation of the Quantum Fisher information}

In this section we present the details of the derivation of Eq.~(6). 
The full expression for the QFI is
\begin{align}
  F_q=2\sum_{i,j|\lambda_i+\lambda_j\neq0}\frac{\modsq{\bra{\psi_i}\partial_\theta\hat\varrho^{(rc)}\ket{\psi_j}}}{\lambda_i+\lambda_j}.
\end{align}
The sum can be split into $i=j$ and $i\neq j$, giving
\begin{align}
  F_q=4\modsq{\bra{\psi_1}\dot{\hat\varrho}_{rc}^{(\theta)}(t)\ket{\psi_2}}+\frac1{\lambda_1}\modsq{\bra{\psi_1}\dot{\hat\varrho}_{rc}^{(\theta)}(t)\ket{\psi_1}}
  +\frac1{\lambda_2}\modsq{\bra{\psi_2}\dot{\hat\varrho}_{rc}^{(\theta)}(t)\ket{\psi_2}}.    
\end{align}
We now separately focus on the first term of this sum, which is
By writing
\begin{align}
  \bra{\psi_1}\dot{\hat\varrho}_{rc}^{(\theta)}(t)\ket{\psi_2}=ae^{i\phi},\ \ \ a\in\mathbb R,
\end{align}
we notice that 
\begin{align}
  \modsq{\bra{\psi_1}\dot{\hat\varrho}_{rc}^{(\theta)}(t)\ket{\psi_2}}=a^2=\frac14\modsq{\trr{\dot{\hat\varrho}_{rc}^{(\theta)}(t)\Big[(\ketbra{\psi_1}{\psi_2}+\ketbra{\psi_2}{\psi_1})\cos\phi
      +\frac1i(\ketbra{\psi_1}{\psi_2}-\ketbra{\psi_2}{\psi_1})\sin\phi\Big]}}
\end{align}
Hence  by introducing
\begin{align}
  \hat\sigma_j=\frac1{\sqrt{\lambda_j}}\ketbra{\psi_j}{\psi_j},\ \ \ j=1,2,\ \ \ \hat\sigma_3=\hat\sigma_{rc}^{(x)}\cos\phi+\hat\sigma_{rc}^{(y)}\sin\phi,
\end{align}
with
\begin{align}
  \hat\sigma_{rc}^{(x)}=\ketbra{\psi_1}{\psi_2}+\ketbra{\psi_2}{\psi_1},\ \ \ \hat\sigma_{rc}^{(y)}=\frac1i(\ketbra{\psi_1}{\psi_2}-\ketbra{\psi_2}{\psi_1})
\end{align}
we obtain
\begin{align}
  \I=\sum_{i=1}^3\modsq{\tr{\hat\sigma_i\dot{\hat\varrho}(\theta,t)}},
\end{align}
as reported in the main text.

\setcounter{equation}{0}
\renewcommand{\theequation}{B.\arabic{equation}}

\section{The QFI for the XY model}

The initial source state
\begin{align}
  \ket{\psi}=\ket{0}_{sr}
\end{align}
evolves under the local transformation $e^{-i\theta\hat\sigma_{sr}^{(x)}}$ to
\begin{align}\label{eq.st.init}
  \ket{0}_{sr}\longrightarrow\cos\theta\ket0_{sr}-i\sin\theta\ket{1}_{sr}.
\end{align}
Hence the state that undergoes the time evolution is (we label the source qubit with index 0 and use $\alpha=\cos\theta$, $\beta=-i\sin\theta$)
\begin{align}
  \ket{\psi(\theta;t=0)}=\alpha\0+\beta\ket{1}_0\bigotimes_{\kappa\neq 0}\ket0_\kappa,
\end{align}
and it is propagated with the Hamiltonian from Eq.~(11).  
The first part, $\0$, does not evolve under the Hamiltonian~(11) (it is its eigenstate with zero eigenvalue), while the evolution of the second, i.e, 
\begin{align}
  \ket\phi=\ket{1}_0\bigotimes_{\kappa\neq 0}\ket0_\kappa
\end{align}
can be found using the Taylor series of the evolution operator. The one-fold action of $\hat H$ on
the state gives 
\begin{align}
  \hat H\ket\phi=\ket1_1\bigotimes_{\kappa\neq1}\ket0_\kappa+\ket1_{-1}\bigotimes_{\kappa\neq-1}\ket0_\kappa.
\end{align}
Both the components of the resulting state have the same form as the initial one, but now the excitations have moved to the adjacent qubits. Another action of the Hamiltonian yields
\begin{align}
  \hat H^2\ket{\phi}=\ket1_2\bigotimes_{\kappa\neq2}\ket0_\kappa+2\ket1_{0}\bigotimes_{\kappa\neq0}\ket0_\kappa+\ket1_{-2}\bigotimes_{\kappa\neq-2}\ket0_\kappa.
\end{align}
In general, the $\mu$-fold action will give the coefficients of the Newton binomial and the excitation of odd/even kets, depending on the parity of $\mu$. 
\begin{figure}[t!]
    \centering
    \includegraphics[width=0.4\linewidth]{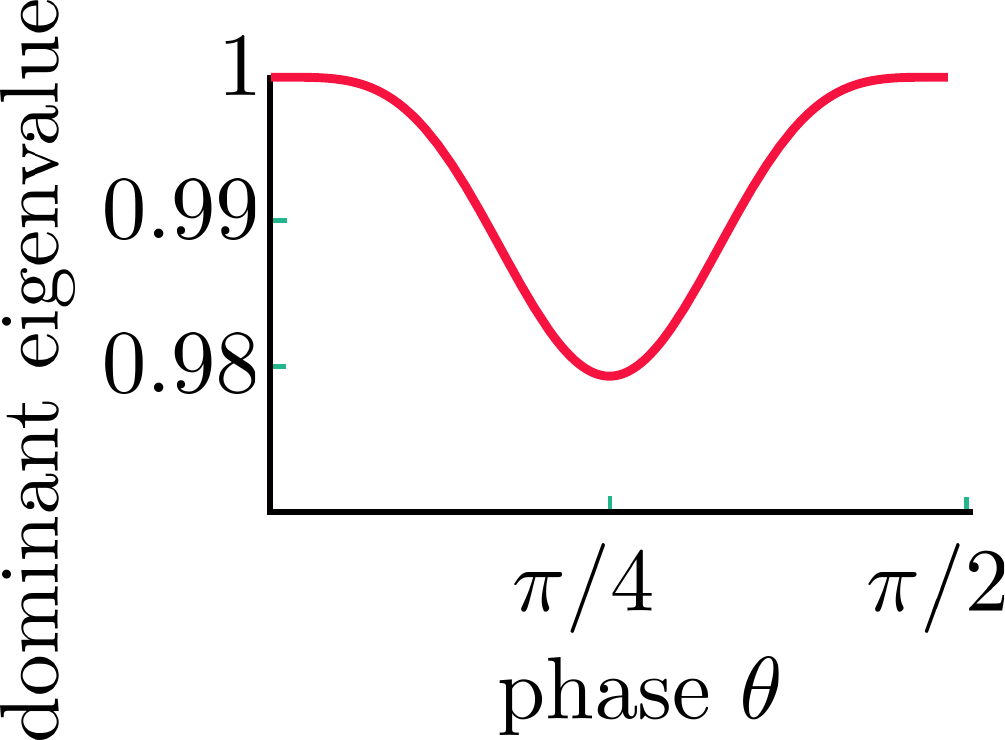}
    \caption{The dominant eigenvalue of the receiver density matrix from Eq.~\eqref{eq.app.rc} with $n=100$ and $t=52$.}
    \label{fig.app.eig}
\end{figure}

Thus the state will take the form
\begin{align}
  \hat H^\mu\ket{\phi}=\sum_{\nu=0}^\mu\binom \mu\nu\ket1_{\kappa_\nu}\bigotimes_{\kappa\neq\kappa_\nu}\ket0_\kappa,
\end{align}
where $\kappa_\nu=2\nu-\mu$. The full state at time $t$ is
\begin{align}\label{eq.st.t}
  \ket{\psi(\theta;t)}=\alpha\0+\beta\sum_{\mu=1}^\infty\frac{(-it)^\mu}{\mu!}\sum_{\nu=0}^\mu\binom \mu\nu\ket1_{\kappa_\nu}\bigotimes_{\kappa\neq\kappa_\nu}\ket0_\kappa.
\end{align}
The density matrix for this pure state is
\begin{align}
  \hat\varrho(\theta;t)=\ketbra{\psi(\theta;t)}{\psi(\theta;t)}
\end{align}
and according to Eq.~(4), the state of the receiver is obtained by tracing out the non-receiver degrees of freedom, i.e., 
\begin{align}
  \hat\varrho_{rc}^{(\theta)}(t)=\titr{\hat\varrho(\theta;t)}.
\end{align}
After some algebraic manipulations we obtain
\begin{align}\label{eq.app.rc}
  \hat\varrho_{rc}^{(\theta)}=\ketbra00_{rc}\left(1-\modsq\beta J^2_n(2t)\right)+\alpha J_n(2t)(\ketbra01_{rc}\beta^*+\ketbra10_{rc}\beta)+\modsq\beta J^2_n(2t)\ketbra11_{rc}.
\end{align}
The dominant eigenvalue is shown in Fig.~\ref{fig.app.eig} using exemplary values of $n=100$ and $t=52$. The state is pure only for $\theta=0$.
Expanding for small $\theta$ around 0 we obtain
\begin{align}
  \hat\varrho_{rc}^{(\theta)}=\ketbra00_{rc}(1-\theta^2J^2_n(2t))+i\theta J_n(2t)(\ketbra01_{rc}-\ketbra10_{rc})+\theta^2J^2_n(2t)\ketbra11_{rc} ,
\end{align}
which is, up to the dominant terms in small $\theta$, the density matrix of the pure state from Eq.~(14) reported in the main text.

\end{document}